\documentclass[aps,pre,twocolumn,showpacs,preprintnumbers,a4paper]{revtex4} 

\usepackage{graphics}
\usepackage{epsf}
\usepackage{eufrak}
\usepackage{epsfig}

\begin{document} 

\flushbottom

\title{Violation of the Widom scaling law for effective crossover exponents}

\author{S. L\"ubeck}
\affiliation{Weizmann Institute of Science, 
Department of Physics of Complex Systems, 
76100 Rehovot, Israel,\\
Theoretische Tieftemperaturphysik,
Gerhard-Mercator-Universit\"at, 
47048 Duisburg, Germany}

\date{August 6, 2003}

\begin{abstract}
In this work we consider the universal crossover behavior 
of two non-equilibrium systems exhibiting a continuous 
phase transition.
Focusing on the field driven crossover 
from mean-field to non-mean-field scaling behavior 
we show that the well-known Widom scaling law
is violated for the effective exponents in the so-called
crossover regime.
\end{abstract}



\preprint{accepted for publication in {\it Physical Review E} 2004}

\maketitle

\section{Introduction}

A foundation for the understanding of critical phenomena
was provided by Wilson's renormalization group (RG) 
approach~\cite{WILSON_1,WILSON_2} 
which maps the critical point onto a fixed point of a
certain transformation of the system's Hamiltonian,
Langevin equation, etc.
The RG theory presents a powerful tool to 
estimate the values of the critical exponents,
it allows to predict which parameters 
determine the universality class
and it explains the existence of an 
upper critical dimension~$D_{\text c}$ above which the 
mean-field theory applies.  
Furthermore, crossover phenomena between two different universality 
classes are well understood in terms
of competing fixpoints~(see for instance~\cite{YEOMANS_1}).
Nevertheless there are still some open aspects of 
crossover phenomena which are discussed in the
literature.
The question whether the crossover 
scaling functions are universal was revisited
several
times~\cite{LUIJTEN_1,LUIJTEN_3,LUIJTEN_2,PELISSETTO_1,PELISSETTO_2,CARACCIOLO_1,LUEB_29}.
For instance it was shown recentlty
that two different models, belonging to
the same universality class, are characterized by the same (universal) 
crossover scaling functions~\cite{LUEB_29}.
This is remarkable, since the universal scaling behavior is usually restricted to
a small vicinity around the critical point.
In the case of a crossover the universal scaling
functions span several decades in temperature or 
conjugated field.

Another question of interest concerns the so-called effective
exponents~\cite{RIEDEL_1} which can be defined as logarithmic 
derivatives of the corresponding scaling functions.
It is still open whether these effective exponents
fulfill over the full crossover the well-known 
scaling laws which connect critical exponents.
This question is closely related to the more general and 
very important question whether effective exponents
obey the scaling laws at all.
For instance it is known experimentally~\cite{GREYWALL_1} 
as well as theoretically~\cite{WEGNER_3}
that the asymptotic scaling behavior
is often masked by corrections to scaling, so-called confluent 
singularities.
In this case it is useful to analyze the data
in terms of effective exponents and the 
above question naturally arises~\cite{AHLERS_1}.
Thus the validity of the scaling laws
for effective exponents was addressed in experimental
works, RG approaches as well as numerical simulations.
In particular the violation of the
scaling laws for effective exponents
was conjectured from a RG approximation~\cite{CHANG_1}.
But neither experimental nor numerical work could clearly
confirm this conjecture so far.
For instance Binder and Luijten considered numerically a 
crossover in the Ising model and discussed the validity of 
the Rushbrook scaling law~\cite{LUIJTEN_2}.
The observed nonmonotonic crossover
behavior suggests again a violation of the 
Rushbrook scaling law.
However it can not be considered as a rigorous proof.

In this work we consider a field driven crossover
(in the so-called critical crossover limit~\cite{LUIJTEN_1,PELISSETTO_1})
in two different models exhibiting a non-equilibrium
second order phase transition.
Varying the range of interactions we investigate
the crossover from mean-field to non-mean-field
scaling behavior.
The order parameter and the  
order parameter susceptibility is measured as a function
of the conjugated field and
we are able to determine the
corresponding effective exponents over the 
full crossover regime.
Our results show
that the well-known Widom scaling law
is clearly violated for the effective
crossover exponents.
Furthermore we present a simple analytical argument,
suggesting that the scaling laws are valid for 
the asymptotic scaling regimes (where the systems are characterized
by a pure algebraic behavior), whereas
the scaling laws do not hold for the crossover regime (characterized
by a non algebraic behavior).

\section{Models and Simulations}

In the following we consider two different cellular automata
exhibiting a so-called absorbing phase transition.
The first model is the conserved transfer
threshold process (CTTP)~\cite{ROSSI_1}.
In this model lattice sites may be empty, 
occupied by one particle,
or occupied by two particles.
Empty and single occupied sites are considered as
inactive whereas double occupied lattice sites are 
considered as active.
In the latter case one tries to transfer both particles
of a given active site to randomly chosen empty or single
occupied nearest neighbor sites.

The second model is a modified version of the
Manna sandpile model~\cite{MANNA_2}, the fixed-energy
Manna model~\cite{VESPIGNANI_4}.
In contrast to the CTTP the Manna model allows unlimited
particle occupation of lattice sites.
Lattice sites which are occupied by at least two particles
are considered as active and all 
particles are moved to the neighboring
sites selected at random.

In our simulations (see~\cite{LUEB_22,LUEB_24} for 
details) we have used square lattices (with periodic
boundaries) of linear size 
$L\leq 2048$. 
All simulations start from a random distribution
of particles.
After a transient regime both models reach a steady state 
characterized by the density of active sites~$\rho_{\scriptscriptstyle \text a}$.
The density~$\rho_{\scriptscriptstyle \text a}$ is the order parameter
and the particle density~$\rho$ is the control parameter
of the absorbing phase transition, i.e., the order parameter
vanishes at the critical density~$\rho_{\text c}$ according to,
$\rho_{\scriptscriptstyle \text a} \propto \delta\rho^{\beta}$ ,
with the reduced control parameter 
$\delta\rho=\rho/\rho_{\text c}-1$. 
Below the critical density (in the absorbing phase)
the order parameter 
is zero in the steady state.

Similar to equilibrium phase transitions it is 
possible in the case of absorbing phase transitions
to apply an external field~$h$ which is
conjugated to the order parameter.
The conjugated field has to act as
a spontaneous creation of active particles, 
destroying the absorbing state
and therefore the phase transition itself.
Furthermore, the associated linear response function
$\chi_{\scriptscriptstyle \text a}
= {\partial \rho_{\scriptscriptstyle \text a}}/{\partial h}$
has to diverge at the critical point ($\delta\rho=0$, $h=0$).
A realization of the external field 
for absorbing phase transitions with a conserved field
was recently developed in~\cite{LUEB_22} where the
external field triggers movements of inactive
particles which may be activated in this way.
At the critical density~$\rho_{\scriptscriptstyle \text c}$
the order parameter scales as
$\rho_{\scriptscriptstyle \text a} \propto h^{\beta/\sigma} $ .
Using the conjugated field it is possible to investigate
the equation of 
state~$\rho_{\scriptscriptstyle \text a}(\rho,h)$, 
i.e., the order parameter as a function
of both the control parameter and the external field.
A recently performed scaling analysis reveals that
the CTTP and the Manna model are characterized by the
same critical exponents as well as by the same universal
scaling form of the equation of state, i.e., both
models belong to the same universality class~\cite{LUEB_26}.

According to the above definition
particles of active sites are moved to nearest
neighbors only, i.e., the range of interactions 
is $R=1$.
It is straightforward to implement various ranges
of interactions into these models~\cite{LUEB_29}.
In this modified models particles of active sites
are moved (according to the rules of each model) 
to randomly selected sites within a 
radius~$R$. 
Since the dynamics of both considered models is characterized by
simple particle hopping processes, 
various interaction ranges can be easily implemented and 
high accurate data are available.
This is a significant advantage compared to e.g.~equilibrium
system like the Ising model where the increasing 
interaction range causes a slowing down of the dynamics. 

For any finite interaction range the phase transition
is characterized by non-mean-field scaling behavior
which now takes place at the critical 
density~$\rho_{\scriptscriptstyle {\text c}, R}$.
A mean-field phase transition occur for 
infinite interactions ($R \to \infty$) only. 
But mean-field behavior could occur away from the critical
point if the long range interactions reduce the 
critical fluctuations sufficiently.
The crossover between the mean-field and non-mean-field
scaling regimes is described by the famous  
Ginzburg criterion~\cite{GINZBURG_1}
which states that the mean-field picture is self-consistent
in the active phase as long as the fluctuations within a 
correlation volume are small compared 
to the order parameter itself.
This leads for zero field to the crossover condition
${\cal O} [ R_{\scriptscriptstyle \text {eff}} \, 
(\rho-\rho_{\scriptscriptstyle {\text c}, 
R})^{\phi} ] = 1$,
with the crossover exponent $\phi = (4-D)/2 D$~\cite{LUEB_29}.
In order to avoid lattice effects we use
the effective range of interactions~\cite{MON_1}
\begin{equation}
R_{\scriptscriptstyle \text {eff}}^2 \; = \;
\frac{1}{z} \, \sum_{i \neq j}  
| {\underline r}_{\scriptscriptstyle i} \, - \,
{\underline r}_{\scriptscriptstyle j}|^2 \, , \quad
| {\underline r}_{\scriptscriptstyle i} \, - \,
{\underline r}_{\scriptscriptstyle j}| \le R \, ,
\label{eq:def_R_eff}
\end{equation}
where $z$ denotes the number of lattice sites 
within a radius~$R$.

\section{Universal crossover scaling}
\label{sec:uni_scal}

The crossover scaling function 
has to incorporate three scaling fields
(the control parameter, the external field,
and the range of interactions), i.e.,
we make the phenomenological ansatz
\begin{eqnarray}
\label{eq:scal_ansatz_EqoS_co}
&&\rho_{\scriptscriptstyle \text a} 
(\rho, h, R_{\scriptscriptstyle \text {eff}}) 
\;  \sim \;  \\
\,&\,&\lambda^{-\beta_{\scriptscriptstyle \text {MF}}} 
\, \, {\tilde {\EuFrak R}}
({\EuFrak a}_{\scriptscriptstyle \rho} (\rho-\rho_{\scriptscriptstyle {\text c}, R}) 
\; \lambda, 
{\EuFrak a}_{\scriptscriptstyle h} h \; \lambda^{\sigma_{\scriptscriptstyle \text {MF}}} ,
{\EuFrak a}_{\scriptscriptstyle \text R}^{-1} R_{\scriptscriptstyle \text {eff}}^{-1} 
\; \lambda^{\phi} 
) , \nonumber 
\end{eqnarray}
where the universal scaling function ${\tilde {\EuFrak R}}$ 
is the same for all models belonging to a given
universality class whereas all non-universal
system-dependent features (e.g.~the lattice structure,
the update scheme, etc.) are contained in the so-called
non-universal metric factors
${\EuFrak a}_{\scriptscriptstyle \rho}$, 
${\EuFrak a}_{\scriptscriptstyle h}$, 
and ${\EuFrak a}_{\scriptscriptstyle \text R}$~\cite{PRIVMAN_3}.
These factors are determined
by three conditions which normalize 
the scaling function ${\tilde {\EuFrak R}}$.
First, the analytically known 
mean-field scaling function~\cite{LUEB_25,LUEB_26} 
\begin{equation}
{\tilde R}_{\scriptscriptstyle {\text {MF}}}(x,y) 
\; = \; \frac{x}{2} + \sqrt{y+\left (\frac{x}{2} \right )^2}
\label{eq:mf_limit_R}
\end{equation}
should be recovered
for $R\to \infty$, i.e., 
${\tilde {\EuFrak R}}(x,y,0)={\tilde R}_{\scriptscriptstyle {\text {MF}}}(x,y)$.
Therefore,
${\tilde {\EuFrak R}}(1,0,0)  = 
{\tilde R}_{\scriptscriptstyle {\text {MF}}}(1,0) = 1$,
${\tilde {\EuFrak R}}(0,1,0) =  
{\tilde R}_{\scriptscriptstyle {\text {MF}}}(0,1) = 1$,
implying
${\EuFrak a}_{\scriptscriptstyle \rho}  = 
{a_{\scriptscriptstyle \rho, R\to\infty}}/
{\rho_{\scriptscriptstyle {\text c}, R\to\infty}}$
and
${\EuFrak a}_{\scriptscriptstyle h}  = 
{a_{\scriptscriptstyle h, R\to\infty}}$.
Finally, the non-universal metric factor~${\EuFrak a}_{\scriptscriptstyle \text R}$
can be determined by the condition
${\tilde {\EuFrak R}}(x,0,1) \; \sim \;
x^{\beta_{\scriptscriptstyle D}}$ 
for $x\to 0$ yielding~\cite{LUEB_29}
\begin{equation}
{\EuFrak a}_{\scriptscriptstyle \text R} \; = \;
\left ( \frac{\rho_{\scriptscriptstyle {\text c},R=1}}
{a_{\scriptscriptstyle \rho, R=1}}\,
\frac{a_{\scriptscriptstyle \rho,R\to \infty}}
{\rho_{\scriptscriptstyle {\text c},R\to \infty}}
\right )^{{\phi \beta_{\scriptscriptstyle D}}/
{(\beta_{\scriptscriptstyle \text {MF}}-\beta_{\scriptscriptstyle D}})}  .
\label{eq:metric_factors_fractur_R}
\end{equation}
The metric factors were already determined in previous
works~\cite{LUEB_25,LUEB_29}, thus no parameter
fitting is needed in order to perform the following
scaling analysis.

\begin{figure}[t]
  \includegraphics[width=8.0cm,angle=0]{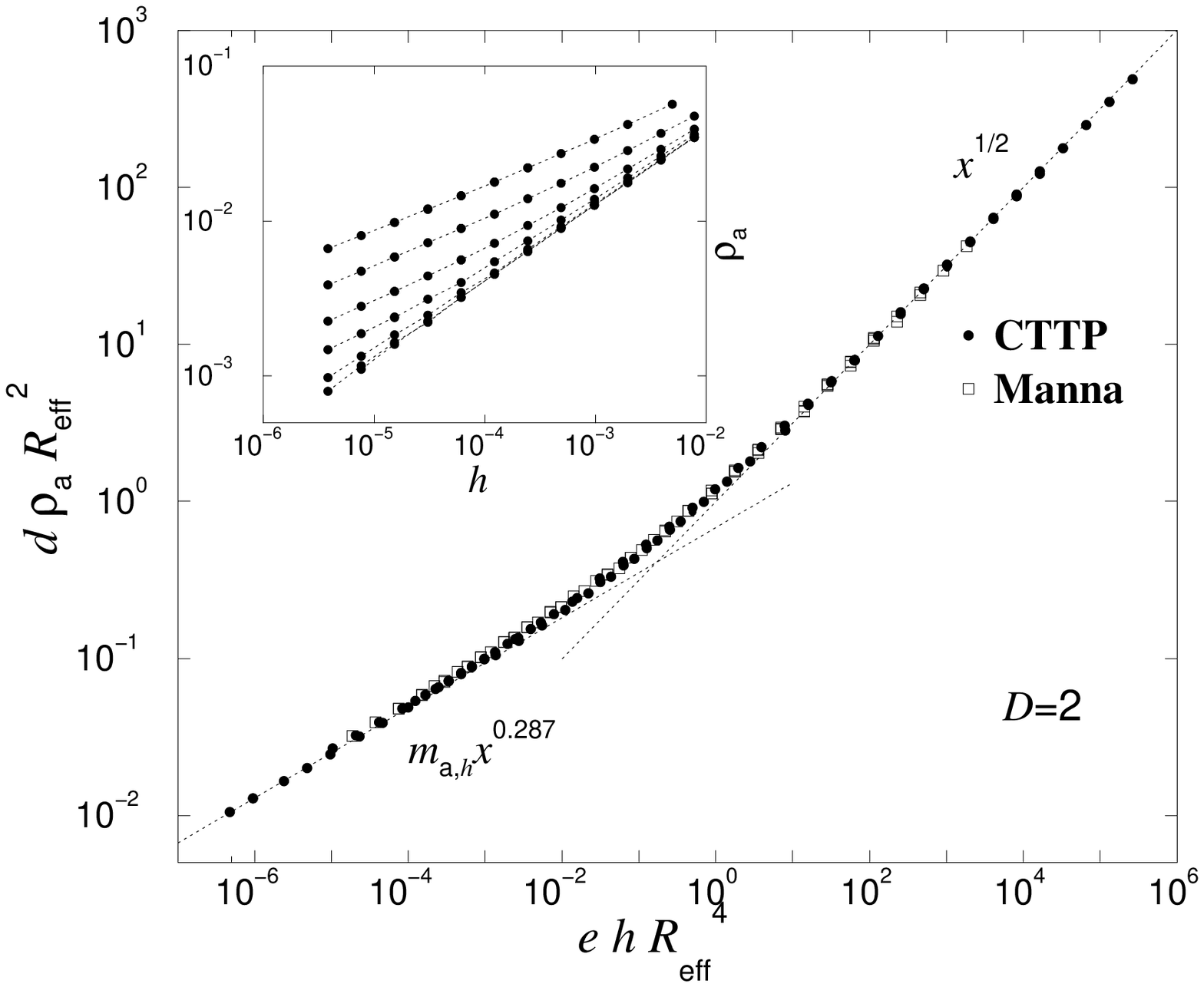}
  \caption{
    The universal crossover scaling function ${\tilde {\EuFrak R}}(0,x,1)$
    of the order parameter
    of the CTTP and Manna model at the critical density for $D=2$.
    The metric factors are given by 
    $e={\EuFrak a}_{\scriptscriptstyle h} 
    {\EuFrak a}_{\scriptscriptstyle \text R}^4$
    and 
    $d={\EuFrak a}_{\scriptscriptstyle \text R}^2$.    
    The dashed lines correspond to the asymptotic behavior
    of the two-dimensional system 
    ($\beta_{\scriptscriptstyle D=2}/\sigma_{\scriptscriptstyle D=2}=0.287$)
    and of the mean-field behavior
    ($\beta_{\scriptscriptstyle \text {MF}}/\sigma_{\scriptscriptstyle \text {MF}}=1/2$).
    The universal amplitude is given 
    by $m_{\scriptscriptstyle {\text a} , h} =0.681$.
    The inset displays the order parameter of the CTTP 
    for various values of the interaction 
    range~$R\in\{1,2,4,..,128\}$ (from top to bottom).
    The dashed lines are just to guide the eyes.
    }
  \label{fig:uni_co_opcp_03} 
\end{figure}

In this work we focus our attention to the field driven crossover,
i.e., we consider the CTTP and the Manna model at the critical
densities $\rho_{\scriptscriptstyle {\text c},R}$
which were determined in~\cite{LUEB_29}.
In Fig.\,\ref{fig:uni_co_opcp_03} we plot the corresponding data  of the 
CTTP for various values of the interaction range~$R$.
As one can see the power law behavior of the order
parameter changes with increasing range of interactions.

The scaling form at the critical point is given by
(setting ${\EuFrak a}_{\scriptscriptstyle \text R}^{-1}
 R_{\scriptscriptstyle \text {eff}}^{-1}\, \lambda^\phi=1$)
\begin{eqnarray}
\label{eq:scal_plot_OPcp_co}   
&&\rho_{\scriptscriptstyle \text a} 
(\rho_{\scriptscriptstyle \text {c},R}, h, R_{\scriptscriptstyle \text {eff}}) 
\;  \sim \;  \\
\,&\,&({\EuFrak a}_{\scriptscriptstyle \text R} 
R_{\scriptscriptstyle \text {eff}})^{-\beta_{\scriptscriptstyle \text {MF}}/\phi} 
\, \, {\tilde {\EuFrak R}}
(0,{\EuFrak a}_{\scriptscriptstyle h} h 
\;
{\EuFrak a}_{\scriptscriptstyle \text R}^{\sigma_{\scriptscriptstyle \text {MF}}/\phi} 
R_{\scriptscriptstyle \text {eff}}^{\sigma_{\scriptscriptstyle \text {MF}}/\phi} 
,1) \nonumber   ,
\end{eqnarray}
with $\beta_{\scriptscriptstyle {\text {MF}}}=1$ and 
$\sigma_{\scriptscriptstyle {\text {MF}}}=2$, respectively.
For sufficiently small field the universal function scales as
\begin{equation}
{\tilde {\EuFrak R}}(0,x,1) \; \sim \; m_{\scriptscriptstyle {\text a} ,h} \,
x^{\beta_{\scriptscriptstyle D}/\sigma_{\scriptscriptstyle D}} \, ,
\quad\quad {\text {for}} \quad x\to 0 \, , 
\label{eq:fractur_R_small_x_field}
\end{equation}
with the universal amplitude $m_{\scriptscriptstyle {\text a} ,h}$.
The scaling form Eq.\,(\ref{eq:scal_plot_OPcp_co})
has to equal for $R=1$ the $D$-dimensional
scaling ansatz 
[ $\rho_{\scriptscriptstyle {\text a}} \sim
(a_{\scriptscriptstyle h, R=1} 
h)^{\beta_{\scriptscriptstyle D}/\sigma_{\scriptscriptstyle D}}$ ]
leading to 
\begin{equation}
\label{eq:uni_ampl_m_h}
m_{\scriptscriptstyle {\text a} , h} \; =
\left ( \frac{a_{\scriptscriptstyle h, R=1}}
{a_{\scriptscriptstyle h, R\to \infty}} 
\right )^{\beta_{\scriptscriptstyle D}/\sigma_{\scriptscriptstyle D}} \;
{\EuFrak a}_{\scriptscriptstyle R}^{\beta_{\scriptscriptstyle {\text {MF}}}/\phi
- \sigma_{\scriptscriptstyle {\text {MF}}} \beta_{\scriptscriptstyle D}/
\sigma_{\scriptscriptstyle D} \phi}\, .
\end{equation}
According to the scaling form Eq.\,(\ref{eq:scal_plot_OPcp_co})
we plot in Fig.\,\ref{fig:uni_co_opcp_03} the
rescaled order parameter $\rho_a\,({\EuFrak a}_{\scriptscriptstyle \text R} 
R_{\scriptscriptstyle \text {eff}})^{2}$ 
as a function of the rescaled field 
${\EuFrak a}_{\scriptscriptstyle h} h ({\EuFrak a}_{\scriptscriptstyle \text R} 
R_{\scriptscriptstyle \text {eff}})^{4}$.
We observe an excellent data collapse for the full
crossover behavior confirming the above 
phenomenological scaling ansatz.

\begin{figure}[t]
  \includegraphics[width=8.0cm,angle=0]{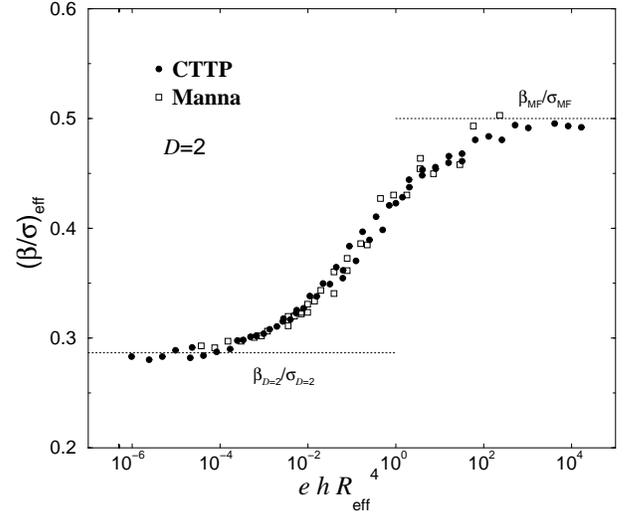}
  \caption{
    The effective exponent~$(\beta/\sigma)_{\scriptscriptstyle \text {eff}}$.  
    }
  \label{fig:uni_co_beta_eff} 
\end{figure}

However, since the entire crossover region covered several 
decades it could be difficult to observe small but systematic
differences between the scaling functions of both models.
Therefore, it its customary to scrutinize  
the crossover via the so-called effective 
exponent~\cite{RIEDEL_1,LUIJTEN_2,CARACCIOLO_1,LUEB_29}
\begin{equation}
\left (
\frac{\beta}{\sigma} 
\right )_{\scriptscriptstyle \text {eff}} \; = \;
\frac{\partial\hphantom{\ln{x}}}{\partial \ln{x}}
\, \ln{{\tilde{\EuFrak R}}(0,x,1)} \, .
\label{eq:def_eff_exp_beta_sigma}
\end{equation}
The perfect collapse of the corresponding data is
shown in Fig.\,\ref{fig:uni_co_beta_eff} and 
confirms again the universality of the crossover
scaling function~${\tilde {\EuFrak R}}$.

Next we consider the order parameter susceptibility.
The scaling form of the susceptibility is given
by
\begin{eqnarray}
\label{eq:scal_ansatz_Sucp_co}   
&& 
{\EuFrak a}_{\scriptscriptstyle \chi} \,
\chi_{\scriptscriptstyle \text a}
(\rho, h, R_{\scriptscriptstyle \text {eff}}) 
\; \sim \; \\
&&\lambda^{\gamma_{\scriptscriptstyle \text {MF}}}\, \, {\tilde {\EuFrak C}}
({\EuFrak a}_{\scriptscriptstyle \rho} (\rho-\rho_{\scriptscriptstyle {\text c}, R})
\; \lambda, 
{\EuFrak a}_{\scriptscriptstyle h} h \; \lambda^{\sigma_{\scriptscriptstyle \text {MF}}},
{\EuFrak a}_{\scriptscriptstyle \text R}^{-1} R_{\scriptscriptstyle \text {eff}}^{-1} 
\; \lambda^{\phi} 
) \, . \nonumber 
\end{eqnarray}
On the other hand the susceptibility is defined
as the derivative of the order parameter with
respect to the conjugated field 
\begin{eqnarray}
\label{eq:def_suscept}
& &\chi_{\scriptscriptstyle \text a}(\rho,h,R_{\scriptscriptstyle \text {eff}}) 
\; = \;
\frac{\partial\hphantom{h}}{\partial h} \,
\rho_{\text a}(\delta\rho, h)  \\ 
& & \sim {\EuFrak a}_{\scriptscriptstyle h} 
\lambda^{\sigma_{\scriptscriptstyle \text {MF}}
-\beta_{\scriptscriptstyle \text {MF}}}
 {\tilde {\EuFrak R}}^{\prime}
({\EuFrak a}_{\scriptscriptstyle \rho} (\rho-\rho_{\scriptscriptstyle {\text c}, R})
\lambda, 
{\EuFrak a}_{\scriptscriptstyle h} h  \lambda^{\sigma_{\scriptscriptstyle \text {MF}}},
{\EuFrak a}_{\scriptscriptstyle \text R}^{-1} R_{\scriptscriptstyle \text {eff}}^{-1} 
\lambda^{\phi} 
) \nonumber 
\end{eqnarray}
with ${\tilde {\EuFrak R}}^{\prime}(x,y,z)=\partial_y {\tilde {\EuFrak R}}(x,y,z)$.
By comparing this expression with Eq.\,(\ref{eq:scal_ansatz_Sucp_co})
we find 
${\tilde {\EuFrak C}}(x,y,z)=\partial_y {\tilde {\EuFrak R}}(x,y,z)$,
${\EuFrak a}_{\scriptscriptstyle \chi}= {\EuFrak a}_{\scriptscriptstyle h}^{-1}$,
as well as the Widom scaling law
\begin{equation}
\gamma \; = \; \sigma \, - \,  \beta
\label{eq:widom}
\end{equation}
which is well known from equilibrium phase transitions.

Again, the mean-field behavior is recovered for $R\to \infty$, i.e.,
${\tilde {\EuFrak C}}(x,y,0) = 
{\tilde C}_{\scriptscriptstyle {\text {MF}}}(x,y) 
= {1}/{2}\; ({y + \left( {x}/{2}\right )^2})^{-1/2}$, 
implying
${\tilde {\EuFrak C}}(1,0,0) = 
{\tilde C}_{\scriptscriptstyle {\text {MF}}}(1,0) = 1$,
${\tilde {\EuFrak C}}(0,1,0) =  
{\tilde C}_{\scriptscriptstyle {\text {MF}}}(0,1) =  {1}/{2}$,
as well as $\gamma_{\scriptscriptstyle {\text {MF}}}=1$.

\begin{figure}[t]
  \includegraphics[width=8.0cm,angle=0]{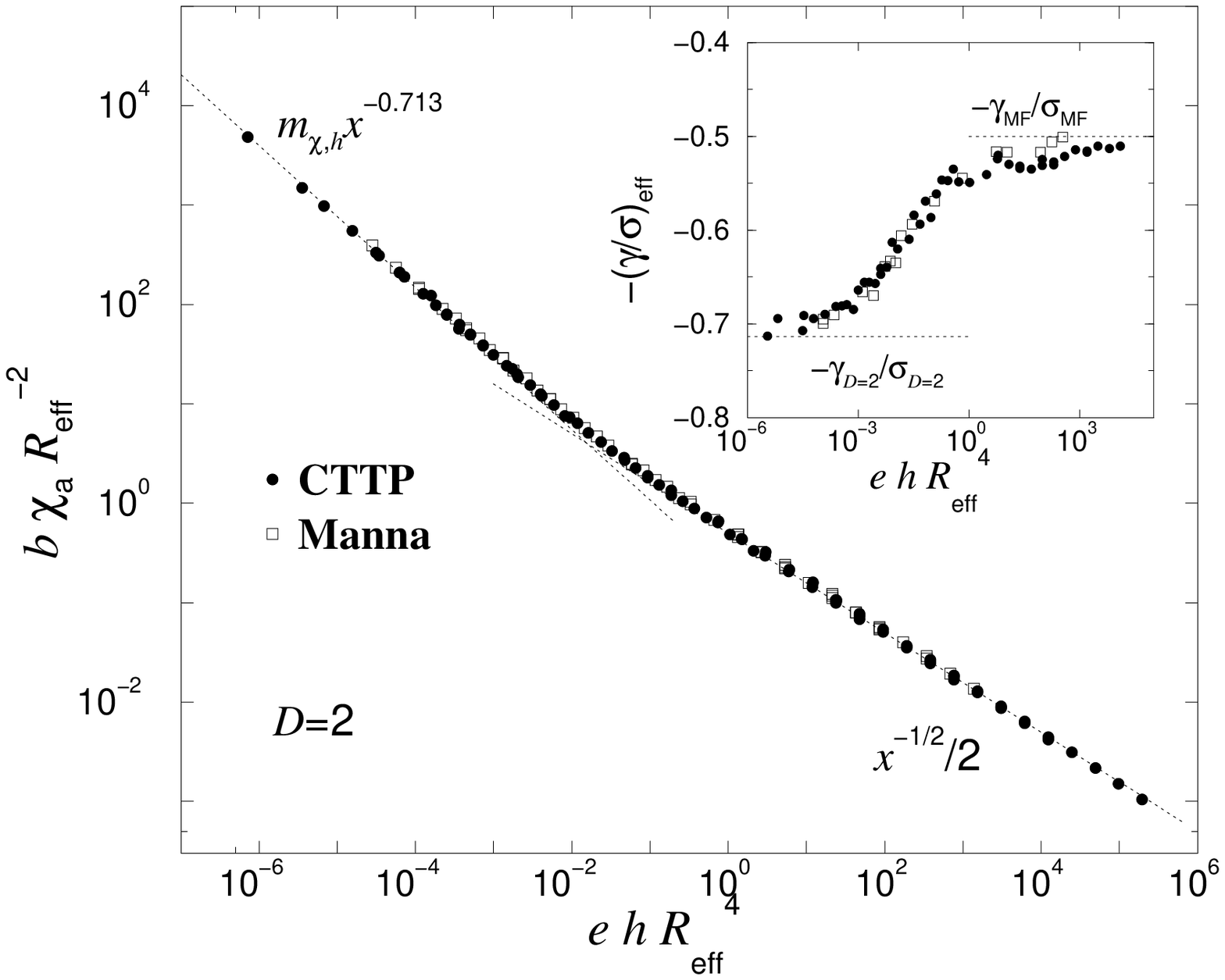}
  \caption{
    The universal crossover scaling function ${\tilde {\EuFrak C}}(0,x,1)$
    of the susceptibility
    of the CTTP and the Manna model at the critical density for $D=2$.
    The metric factors are given by 
    $e={\EuFrak a}_{\scriptscriptstyle h} 
    {\EuFrak a}_{\scriptscriptstyle \text R}^4$
    and 
    $b={\EuFrak a}_{\scriptscriptstyle h}^{-1}
    {\EuFrak a}_{\scriptscriptstyle \text R}^2$.    
    The dashed lines correspond to the asymptotic behavior
    of the two-dimensional system 
    ($\gamma_{\scriptscriptstyle D=2}/\sigma_{\scriptscriptstyle D=2}=0.713$)
    and of the mean-field behavior
    ($\gamma_{\scriptscriptstyle \text {MF}}/\sigma_{\scriptscriptstyle \text {MF}}=1/2$).
    The universal amplitude is given 
    by $m_{\scriptscriptstyle \chi , h} =0.208$.
    The inset displays the corresponding effective
    exponent~$(\gamma/\sigma)_{\scriptscriptstyle \text {eff}}$.  
    }
  \label{fig:uni_co_sucp_02} 
\end{figure}

Similar to the order parameter we plot the susceptibility 
according to the scaling form
\begin{eqnarray} 
\label{eq:scal_plot_Sucp_co}   
&& 
{\EuFrak a}_{\scriptscriptstyle h}^{-1} \,
\chi_{\scriptscriptstyle \text a}
(\rho_{\scriptscriptstyle {\text c}, R}, h, R_{\scriptscriptstyle \text {eff}}) 
\; \sim \; \\
&&
({\EuFrak a}_{\scriptscriptstyle \text R} 
R_{\scriptscriptstyle \text {eff}})^{\gamma_{\scriptscriptstyle \text {MF}}/\phi} 
\, {\tilde {\EuFrak C}}
(0,{\EuFrak a}_{\scriptscriptstyle h} h \; 
({\EuFrak a}_{\scriptscriptstyle \text R} 
R_{\scriptscriptstyle \text {eff}})^{\sigma_{\scriptscriptstyle \text {MF}}/\phi}
,1) \, .\nonumber 
\end{eqnarray}
Approaching the transition point the susceptibility is expected
to scales as
${\tilde {\EuFrak C}}(0,x,1) \sim  m_{\scriptscriptstyle \chi,h} 
x^{-\gamma_{\scriptscriptstyle D}/\sigma_{\scriptscriptstyle D}}$,
for $x \to 0$ , 
where the universal power-law amplitude is given by
\begin{eqnarray}
\label{eq:uni_ampl_m_chi_h}   
&&m_{\scriptscriptstyle \chi,h} \; = \\
&&
\left ( \frac{a_{\scriptscriptstyle h, R=1}}
{a_{\scriptscriptstyle h, R\to \infty}} 
\right )^{1-\gamma_{\scriptscriptstyle D}/\sigma_{\scriptscriptstyle D}} \;
{\EuFrak a}_{\scriptscriptstyle R}^{-\gamma_{\scriptscriptstyle {\text {MF}}}/\phi
+\sigma_{\scriptscriptstyle {\text {MF}}} \gamma_{\scriptscriptstyle D}/
\sigma_{\scriptscriptstyle D} \phi}
\; \frac{\beta_{\scriptscriptstyle D}}{\sigma_{\scriptscriptstyle D}} 
\nonumber .
\end{eqnarray}

The rescaled susceptibility is 
shown in Fig.\,\ref{fig:uni_co_sucp_02}.
Over the entire crossover region we got an excellent
data collapse including both asymptotic scaling regimes.
The inset displays the effective exponent
\begin{equation}
\left (
\frac{\gamma}{\sigma} 
\right )_{\scriptscriptstyle \text {eff}} \; = \; - \,
\frac{\partial\hphantom{\ln{x}}}{\partial \ln{x}}
\, \ln{{\tilde{\EuFrak C}}(0,x,1)} 
\label{eq:def_eff_exp_gamma_sigma}
\end{equation}
which exhibits again a monotonic crossover from the
two-dimensional scaling regime to the mean-field
scaling behavior.

\section{Widom scaling law}

\begin{figure}[t]
  \includegraphics[width=8.0cm,angle=0]{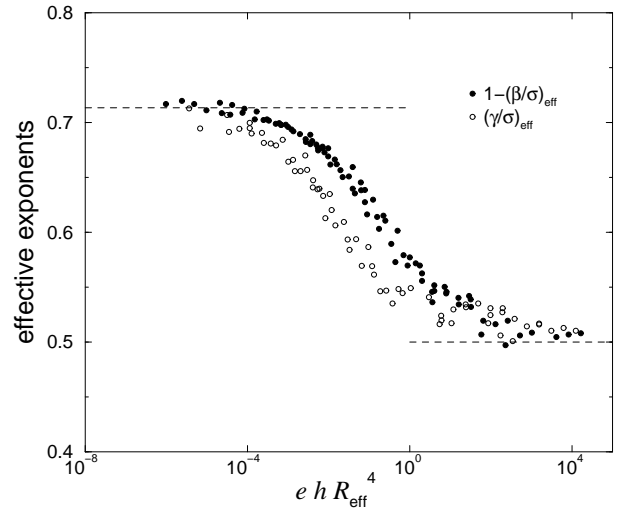}
  \caption{
    The violation of Widom scaling law 
    [Eq.\,(\protect\ref{eq:widom_eff})]
    in the crossover regime.
    }
  \label{fig:widom_01} 
\end{figure}

In this way we have obtained the effective
exponents 
$(\beta / \sigma )_{\scriptscriptstyle \text {eff}}$
and $(\gamma / \sigma )_{\scriptscriptstyle \text {eff}}$
for the field driven crossover from 
mean-field to non-mean-field behavior.
Thus we are able to check the corresponding 
Widom scaling law 
\begin{equation}
\left (
\frac{\gamma}{\sigma}
\right )_{\scriptscriptstyle \text {eff}} \; = \;
1 \, - \, \left (
\frac{\beta}{\sigma}
\right )_{\scriptscriptstyle \text {eff}} \, ,
\label{eq:widom_eff}
\end{equation}
for the whole crossover region.
The corresponding data are shown in Fig.\,\ref{fig:widom_01}.
As can be seen the Widom scaling law is 
fulfilled for the asymptotic regimes ($D=2$ scaling
behavior and mean-field scaling) but it is clearly
violated for the intermediate crossover region.
This result is not surprising if one notices that the
above Widom law [Eq.\,(\ref{eq:widom_eff})] 
corresponds to the differential 
equation [see Eqs.\,(\ref{eq:def_eff_exp_beta_sigma},
\ref{eq:def_eff_exp_gamma_sigma})]
\begin{equation}
- \frac{\partial\ln{\hphantom{x}}}{\partial \ln{x}}
\, \frac{\partial\hphantom{x}}{\partial x}
\, {\tilde{\EuFrak R}}(0,x,1)
\; = \;
1 \, - \,
\frac{\partial\ln{\hphantom{x}}}{\partial \ln{x}}
{\tilde{\EuFrak R}}(0,x,1) \, .
\label{eq:widom_dlg}
\end{equation}
Using $1= \partial\ln{a x} / \partial\ln{x}$ we get
\begin{equation}
- \ln{\partial_{\scriptscriptstyle x} {\tilde{\EuFrak R}}(0,x,1)}
\; = \; \ln{ax} \, - \,
\ln{{\tilde{\EuFrak R}}(0,x,1)} \, + \, c \, ,
\label{eq:widom_dlg_2}
\end{equation}
where $c$ is some constant.
It is straightforward to show that this differential
equation is solved by simple power-laws 
[${\tilde{\EuFrak R}}(0,x,1)= c_{\scriptscriptstyle 0}
x^{c_{\scriptscriptstyle 1}}$
with $c_{\scriptscriptstyle 1}=1/a \exp{c}$].
Thus the Widom scaling law is fulfilled in the asymptotic regimes only.
In the case that the scaling behavior is affected by
crossovers, confluent singularities, etc.~no pure power-laws
occur and the scaling laws do not hold for the corresponding
effective exponents.

\section{Conclusion}

In conclusion, the crossover from mean-field to non-mean-field
scaling behavior is numerically investigated for two 
different models exhibiting a second order phase transition.
Increasing the range of interactions we are able 
to cover the full crossover
region which spans several decades of the conjugated field.
The corresponding data show that the Widom scaling law is
violated in the crossover regime. 
Notice that we focus in our investigations on the 
particular universality class of absorbing phase transitions 
only for technical reasons.
The demonstrated violation of the Widom scaling 
can be applied to continuous phase
transitions in general.


We would like to thank A.~Hucht, K.\,D.~Usadel, and B.~Schnurr 
for helpful discussions.
This work was financially supported by the 
Minerva Foundation (Max Planck Gesellschaft).

\end{document}